\documentclass[12pt,fleqn]{article}
\newcommand{\vek}[1]{\mbox{\bf #1}}
\newcommand{\be}{\begin{equation}}
\newcommand{\ee}{\end{equation}}

\begin{document}
\begin{center}
\Large\bf
Which action 
for brane worlds?
\end{center}
\vspace{2ex}
\normalsize \rm
\begin{center}
 Rainer Dick\\[1ex] 
{\small\it
Department of Physics and Engineering Physics,
University of Saskatchewan\\ 116 Science Place,
Saskatoon, SK S7N 5E2, Canada
}
\end{center}

\vspace{5ex}
\noindent
{\bf Abstract:}
In his pioneering work on singular shells in general relativity,
Lanczos had derived jump conditions across
energy-momentum carrying hypersurfaces
from the Einstein equation with codimension 1 sources.
However, on the level of the action,
the discontinuity of the connection
arising from a codimension 1 energy-momentum source 
requires to take
into account two adjacent space-time regions separated by
the hypersurface. 

The purpose of the present note is to draw attention to the
fact that Lanczos' jump conditions can be derived
from an Einstein action but not from an Einstein--Hilbert
action.

\newpage  
\section{Introduction}

Recently, a particular class of cosmological models 
commonly denoted as brane-world models attracted
a lot of attention. These models are essentially
based
on two assumptions:\\
$\bullet$ Our universe may be described
as a timelike four-dimensional submanifold $\Sigma$ of a
$(1+3+n)$--dimensional bulk space-time ${\cal S}$, with $n\ge 1$
additional spacelike dimensions. \\
$\bullet$ The additional spacelike dimensions can only be probed by
gravity and eventually some non-standard matter
degrees of freedom, but standard model
particles cannot leave our observable universe $\Sigma$.

Predecessors of this kind of cosmological scenarios
rely on dynamical binding mechanisms for low energy matter to
an effectively four-dimensional submanifold, which has
some finite but small extension in the transverse dimensions.
Dynamical mechanisms for explaining such
scenarios have been proposed already
by Akama \cite{akama}, by Rubakov and Shaposhnikov \cite{rubakov},
by Visser \cite{visser},
and by Gibbons and Wiltshire \cite{gibbons},
 and the corresponding transversally
``thick'' universes have also attracted much attention recently
\cite{arkani1}, see
also \cite{tim,grojean} and references there.

The other extreme, which was partly motivated from string theory,
consists of 3-branes $\Sigma$ which have no transverse extension at all
and are strictly codimension-$n$ submanifolds
 \cite{horava,arkani2,burt,pierre1,RS1,keith,klaus}. 
Here the confinement 
of matter to $\Sigma$ is not necessarily 
a dynamical phenomenon, but imposed
axiomatically through the assumption that matter degrees contribute
only a hypersurface integral over $\Sigma$ to an action $S$ which
also contains bulk terms for gravity and eventually a few other bulk
degrees of freedom. Such an axiomatic distinction between
hypersurface and bulk degrees of freedom may seem strange at first
sight, but {\it a priori} there is nothing mathematically inconsistent 
with it, and so there is no {\it a priori} reason to rule out
such scenarios\footnote{Whether or not we find {\it a posteriori}
reasons is another story, of course. But this can only be clarified
by investigations of (post-)Newtonian limits \cite{gata,dimi,wolf1,wolf2,
ioannis}
and of cosmological implications of these models
\cite{pierre1,nihei,groj1,kraus,yuri,pierre2,vollick,ida,cvetic2,
cline,holdom,deruelle,tye}.}.

It has been mentioned already that such $(1+3)$--dimensional
submanifolds go by the name
3-branes, but referring to the old literature
on singular timelike 3-manifolds in
$1+3$ dimensions (e.g.\ \cite{lanczos,israel}) 
another appropriate term would be
hypersurface layers. Layers denote hypersurfaces 
with discontinuous extrinsic curvature across the hypersurface
due to the presence of energy-momentum on the 
hypersurface. 

The purpose of this note is to draw attention to the fact
that an Einstein action instead of an Einstein--Hilbert action
does yield the same jump conditions across a hypersurface
layer as the Einstein equation.

The following notation is used:
\[
\left.f(x)\right|_{x=a}=f(a),
\qquad
\left[f(x)\right]_{x=a}^{x=b}=f(b)-f(a).
\]

It is helpful to start with a toy model from electrodynamics
before we address the Einstein--Hilbert action
in section \ref{gravity1}:

\section{A toy model: Electrodynamics with
planar codimension 1 sources}

Electrodynamics in $1+3$ dimensions 
with codimension 1 sources located on the plane $x^3=0$
is described by an action
\[
S=-\frac{1}{4}\int dt\int d^2\vek{x}\int_{-\infty}^{\infty}dx^3
 F^{\mu\nu}F_{\mu\nu}+\int dt\int d^2\vek{x}\left.(\vek{j}\cdot\vek{A}
+j^0A_0)\right|_{x^3=0},
\]
where $d^2\vek{x}=dx^1dx^2$ and all vectors are two-dimensional
vectors: $\vek{j}\cdot\vek{A}=j^1A_1+j^2A_2$.

Without much ado we write down the equations of motion which
 follow from $\delta S=0$:
\be\label{eq1}
\partial_\mu F^{\mu\nu}=-j^\nu\delta(x^3),
\ee
implying in particular
\be\label{eq2}
\lim_{\epsilon\to +0}
\left[F^{3\nu}\right]_{x^3=-\epsilon}^{x^3=\epsilon}=
-\left. j^\nu\right|_{x^3=0}.
\ee
Of course, this can be confirmed from a more careful evaluation
of the variation of $S$:
\[
S=-\frac{1}{4}\lim_{\epsilon\to +0}\left(\int dt\int d^2\vek{x}
\int_{-\infty}^{-\epsilon} dx^3 F^{\mu\nu}F_{\mu\nu}+
\int dt\int d^2\vek{x}
\int^{\infty}_{\epsilon} dx^3 F^{\mu\nu}F_{\mu\nu}\right)
\]
\[
+\int dt\int d^2\vek{x}\left.(\vek{j}\cdot\vek{A}
+j^0A_0)\right|_{x^3=0},
\]
whence
\[
\delta S=\lim_{\epsilon\to +0}\left(\int dt\int d^2\vek{x}
\int_{-\infty}^{-\epsilon} dx^3\delta
 A_\nu\partial_\mu F^{\mu\nu}
+\int dt\int d^2\vek{x}
\int^{\infty}_{\epsilon} dx^3 
\delta A_\nu\partial_\mu F^{\mu\nu}\right)
\]
\[
-\lim_{\epsilon\to +0}\int dt\int d^2\vek{x}\left[
\delta A_\nu F^{3\nu}\right]^{x^3=-\epsilon}_{x^3=\epsilon}
+\int dt\int d^2\vek{x}\left.
\left(\delta\vek{A}\cdot\vek{j}+\delta A_0 j^0\right)\right|_{x^3=0}.
\]
Therefore, (\ref{eq1}) and the jump condition
(\ref{eq2}) indeed imply $\delta S=0$.

However, this does not work with the
Einstein--Hilbert action:

\section{A first {\sl Ansatz} for the action in
brane models}\label{gravity1}

 For simplicity I pretend that I can cover
a $(1+4)$--dimensional space-time
by a single coordinate patch $x^M$ which is Gaussian
close to the brane:
 $x^0=t$, $x^j$ ($1\le j\le 3$) are tangential to the
world-volume of the brane, and
$x^5$ is normal: $g_{\mu 5}=0$ on the brane,
 $0\le\mu\le 3$. It is known that the geodesic distance
from the brane provides such a coordinate system locally, whence
the brane is localized at $x^5=0$. If this coordinate system
cannot be extended to all of the five-dimensional
 space-time (which is what
one expects), we have to glue together several patches
with appropriate transition functions
to formulate action principles. 
However, the difficulty that we encounter with the Einstein--Hilbert
action is related only to the
boundary conditions across the brane, and therefore we write
the Einstein--Hilbert {\sl Ansatz} for the brane action as
\be\label{action1}
S_{EH}=\int dt\int d^3\vek{x}
\int_{-\infty}^\infty dx^5 \sqrt{-g}\left(\frac{m^3}{2}R-\Lambda\right)
+\left.\int dt\int d^3\vek{x}\,
 {\cal L}\right|_{x^5=0},
\ee
where we assume that the brane Lagrangian ${\cal L}$ contains 
no genuine gravitational terms: Derivatives
of the metric appear
in ${\cal L}$ only through covariant derivatives on fermions
and eventually massive vector fields.

One might expect an Einstein equation
to emerge from (\ref{action1}):
\be\label{einstein1}
R_{MN}-\left(\frac{1}{2}R-\frac{\Lambda}{m^3}\right)g_{MN}
=-\frac{2}{m^3\sqrt{-g}}\delta(x^5)
\frac{\delta{\cal L}}{\delta g^{MN}}.
\ee
However, a naive derivation of (\ref{einstein1}) from
 (\ref{action1}) would have to assume continuity of normal
derivatives across the brane, in {\sl a posteriori} contradiction
to (\ref{einstein1}).

To clarify this and to reveal which equations would really
follow from stationarity of $S_{EH}$, we write it more carefully
as
\be\label{action1new}
S_{EH}=\lim_{\epsilon\to +0}\left(\int dt\int d^3\vek{x}
\int_{-\infty}^{-\epsilon} dx^5
\sqrt{-g}\left(\frac{m^3}{2}R-\Lambda\right)
\right.
\ee
\[
+\left.\int dt\int d^3\vek{x}
\int_{\epsilon}^{\infty} dx^5
\sqrt{-g}\left(\frac{m^3}{2}R-\Lambda\right)
\right)
+\left.\int dt\int d^3\vek{x}\,
 {\cal L}\right|_{x^5=0}.
\]
Variation of the metric then yields
\[
\delta S_{EH}=\frac{m^3}{2}\lim_{\epsilon\to +0}\left(\int dt\int d^3\vek{x}
\int_{-\infty}^{-\epsilon} dx^5
\sqrt{-g}\delta g^{MN}\left(R_{MN}-\frac{1}{2}g_{MN}R
+\frac{\Lambda}{m^3}g_{MN}\right)\right.
\]
\be\label{var1}
+\left.\int dt\int d^3\vek{x}
\int_{\epsilon}^{\infty} dx^5
\sqrt{-g}\delta g^{MN}\left(R_{MN}-\frac{1}{2}g_{MN}R
+\frac{\Lambda}{m^3}g_{MN}\right)\right)
\ee
\[
+\frac{m^3}{2}\lim_{\epsilon\to +0}\int dt\int d^3\vek{x}
\left[\sqrt{-g}\left(g^{MN}\delta\Gamma^5{}_{MN}
-g^{5N}\delta\Gamma^M{}_{MN}\right)\right]^{x^5=-\epsilon}_{x^5=\epsilon}
\]
\[
+\left.\int dt\int d^3\vek{x}\,\delta g^{MN}
\frac{\delta{\cal L}}{\delta g^{MN}}\right|_{x^5=0}
\]
\[
=\frac{m^3}{2}\lim_{\epsilon\to +0}\left(\int dt\int d^3\vek{x}
\int_{-\infty}^{-\epsilon} dx^5
\sqrt{-g}\delta g^{MN}\left(R_{MN}-\frac{1}{2}g_{MN}R
+\frac{\Lambda}{m^3}g_{MN}\right)\right.
\]
\[
+\left.\int dt\int d^3\vek{x}
\int_{\epsilon}^{\infty} dx^5
\sqrt{-g}\delta g^{MN}\left(R_{MN}-\frac{1}{2}g_{MN}R
+\frac{\Lambda}{m^3}g_{MN}\right)\right)
\]
\[
+\frac{m^3}{4}\lim_{\epsilon\to +0}\int dt\int d^3\vek{x}[\sqrt{-g}
(3\delta g^{\mu\nu}g^{55}\partial_5 g_{\mu\nu}
-2\delta g^{\mu 5}g^{55}\partial_\mu g_{55}
\]
\[
-\delta g^{55}g^{\mu\nu}
 \partial_5 g_{\mu\nu}+2g^{55}g_{\mu\nu}\partial_5\delta g^{\mu\nu}
)]^{x^5=-\epsilon}_{x^5=\epsilon}
+\left.\int dt\int d^3\vek{x}\,\delta g^{MN}
\frac{\delta{\cal L}}{\delta g^{MN}}\right|_{x^5=0}.
\]
The jump conditions following from $\delta S_{EH}=0$ are  
incompatible with the
jump conditions following from (\ref{einstein1}) (see
eq.\ (\ref{lanczoseq}) below). 
In fact, $\delta  S_{EH}=0$
would even require a
traceless energy-momentum tensor on the brane
if $\delta{\cal L}/\delta g^{55}=0$. 

In an attempt to infer the jump conditions
 following from (\ref{einstein1}) from an action principle, we
will consider the Einstein action next:
\section{An Einstein action for
brane models}\label{gravity}
The Einstein action
proves more suitable in boundary models \cite{dimi},
and may also be better adapted to brane models:
\[
S_{E}=\lim_{\epsilon\to +0}\left(\int dt\int d^3\vek{x}
\int_{-\infty}^{-\epsilon} dx^5
 \sqrt{-g}\left(\frac{m^3}{2}
g^{MN}\left(\Gamma^K{}_{LM}\Gamma^L{}_{KN}
-\Gamma^K{}_{KL}\Gamma^L{}_{MN}\right)
-\Lambda\right)\right.
\]
\be\label{action2}
+\left.\int dt\int d^3\vek{x}
\int_{\epsilon}^{\infty} dx^5
 \sqrt{-g}\left(\frac{m^3}{2}
g^{MN}\left(\Gamma^K{}_{LM}\Gamma^L{}_{KN}
-\Gamma^K{}_{KL}\Gamma^L{}_{MN}\right)
-\Lambda\right)\right)
\ee
\[
+\left.\int dt\int d^3\vek{x}\,
 {\cal L}\right|_{x^5=0}
\]
\[
=S_{EH}-\frac{m^3}{2}\lim_{\epsilon\to +0}\int dt\int d^3\vek{x}
\left[\sqrt{-g}\left(g^{MN}\Gamma^5{}_{MN}-g^{5N}\Gamma^M{}_{MN}\right)
\right]_{x^5=\epsilon}^{x^5=-\epsilon},
\]
with variation under changes of the metric:
\[
\delta S_{E}=\frac{m^3}{2}\lim_{\epsilon\to +0}\left(\int dt\int d^3\vek{x}
\int_{-\infty}^{-\epsilon} dx^5
\sqrt{-g}\delta g^{MN}\left(R_{MN}-\frac{1}{2}g_{MN}R
+\frac{\Lambda}{m^3}g_{MN}\right)\right.
\]
\be\label{var2}
+\left.\int dt\int d^3\vek{x}
\int_{\epsilon}^{\infty} dx^5
\sqrt{-g}\delta g^{MN}\left(R_{MN}-\frac{1}{2}g_{MN}R
+\frac{\Lambda}{m^3}g_{MN}\right)\right)
\ee
\[
-\frac{m^3}{4}\lim_{\epsilon\to +0}\int dt\int d^3\vek{x}
\left[2\sqrt{-g}\left(\delta g^{MN}\Gamma^5{}_{MN}
-\delta g^{5N}\Gamma^M{}_{MN}\right)\right.
\]
\[
-\left.\sqrt{-g}\delta g^{MN}g_{MN}\left(
g^{KL}\Gamma^5{}_{KL}
- g^{5L}\Gamma^K{}_{KL}\right)\right]^{x^5=-\epsilon}_{x^5=\epsilon}
+\left.\int dt\int d^3\vek{x}\,\delta g^{MN}
\frac{\delta{\cal L}}{\delta g^{MN}}\right|_{x^5=0}
\]
\[
=\frac{m^3}{2}\lim_{\epsilon\to +0}\left(\int dt\int d^3\vek{x}
\int_{-\infty}^{-\epsilon} dx^5
\sqrt{-g}\delta g^{MN}\left(R_{MN}-\frac{1}{2}g_{MN}R
+\frac{\Lambda}{m^3}g_{MN}\right)\right.
\]
\[
+\left.\int dt\int d^3\vek{x}
\int_{\epsilon}^{\infty} dx^5
\sqrt{-g}\delta g^{MN}\left(R_{MN}-\frac{1}{2}g_{MN}R
+\frac{\Lambda}{m^3}g_{MN}\right)\right)
\]
\[
+\frac{m^3}{4}\lim_{\epsilon\to +0}\int dt\int d^3\vek{x}
\left[
\sqrt{-g}\delta g^{\mu\nu}\left(g^{55}\partial_5 g_{\mu\nu}
-g_{\mu\nu}g^{\alpha\beta}g^{55}\partial_5 g_{\alpha\beta}\right)
\right.
\]
\[
-\left.\sqrt{-g}\delta g^{5\mu}\left(g^{55}\partial_\mu g_{55}
-g^{\alpha\beta}\partial_\mu g_{\alpha\beta}\right)
\right]^{x^5=-\epsilon}_{x^5=\epsilon}
+\left.\int dt\int d^3\vek{x}\,\delta g^{MN}
\frac{\delta{\cal L}}{\delta g^{MN}}\right|_{x^5=0}.
\]
$\delta S_E=0$ thus yields 
a five-dimensional Einstein space in the bulk
\[
R_{MN}=\frac{2\Lambda}{3m^3}g_{MN},
\]
and the five-dimensional analog of the Lanczos equations
\cite{lanczos,israel}:
\be\label{lanczoseq}
\lim_{\epsilon\to +0}\left[\partial_5 g_{\mu\nu}
\right]_{x^5=-\epsilon}^{x^5=\epsilon}=
-\left.\frac{2}{m^3}\sqrt{g_{55}}\left(T_{\mu\nu}-\frac{1}{3}
g_{\mu\nu}g^{\alpha\beta}T_{\alpha\beta}\right)\right|_{x^5=0},
\ee
i.e.\ exactly the equations that one infers from
the Einstein equation\footnote{A further equation on the brane 
appears if the brane is a boundary of space-time \cite{dimi}:
In this case the term
$\sim\delta g^{5\mu}$ cannot be cancelled by continuity across
the brane and requires 
\[
\left.g^{55}\partial_\mu g_{55}\right|_{x^5=0}
=\left.g^{\alpha\beta}\partial_\mu g_{\alpha\beta}\right|_{x^5=0},
\]
i.e.\ $g_{55}\sim -\mbox{det}(g_{\alpha\beta})$ on a boundary.} 
(\ref{einstein1}). Here the brane energy-momentum
tensor is defined via
\[
T_{\mu\nu}=-\left.\frac{2}{\sqrt{-\mbox{det}(g_{\alpha\beta})}}
\frac{\delta{\cal L}}{\delta g^{\mu\nu}}\right|_{x^5=0}.
\]
Another advantage of the Einstein action is the
disappearance of $\delta g^{55}$ on the brane,
implying that the Einstein action complies with the usual
assumption that the brane Lagrangian ${\cal L}$ depends
only on the induced metric on the brane.

\section{Conclusion}

The jump conditions following from the Einstein equation
with brane sources imply stationarity of the
Einstein action with brane sources, 
but not of the Einstein--Hilbert action with
brane sources. Furthermore, stationarity
of the Einstein action
complies with brane Lagrangians ${\cal L}$ which do not
depend on the normal component of the metric.

One might be concerned about diffeomorphism invariance
since the Einstein action is only invariant under $IGL(5)$
transformations.
However, we have seen that
stationarity of the Einstein action is equivalent to the
fully covariant Einstein equation (\ref{einstein1}) (remembering
that $x^5$ is a geodesic distance, i.e.\ a well-defined
geometric object). Therefore, besides the
numerical value of the action itself, no classical results
inferred from the use of an Einstein action depend
on the coordinate system.

Note added: An important reference on early 
investigations in brane cosmology is Chamblin and Reall \cite{reall}.
These authors had also already recognized the difficulty with the 
Einstein--Hilbert
action for thin branes and added a Gibbons--Hawking
term to cure the problem.


\begin{thebibliography}{88} 
\bibitem{akama}K.\ Akama, {\sl Pregeometry}, in:
 Gauge Theory and Gravitation, edited by K.\ Kikkawa, N.\ Nakanishi
 and H.\ Nariai, Springer--Verlag, Berlin 1983, pp.\ 267--271.
\bibitem{rubakov}V.A.\ Rubakov,  M.E.\ Shaposhnikov,
  Phys.\ Lett.\ 125B (1983) 136.
\bibitem{visser}M.\ Visser, Phys.\ Lett.\ 159B (1985) 22.
\bibitem{gibbons}G.W.\ Gibbons, D.L.\ Wiltshire,
  Nucl.\ Phys.\ B287 (1987) 717.
\bibitem{arkani1}N.\ Arkani-Hamed, S.\ Dimopoulos, G.\ Dvali,
 Phys.\ Lett.\ B429 (1998) 263.
\bibitem{tim}C.\ Cs\'{a}ki, J.\ Erlich, T.J.\ Hollowood,
 Y.\ Shirman, Nucl.\ Phys.\ B581 (2000) 309. 
\bibitem{grojean}C.\ Grojean, 
  Phys.\ Lett.\ B479 (2000) 273.
\bibitem{horava}P.\ Ho\v{r}ava, E.\ Witten,
 Nucl.\ Phys.\ B460 (1996) 506, B475 (1996) 94.
\bibitem{arkani2}I.\ Antoniadis,
 N.\ Arkani-Hamed, S.\ Dimopoulos, G.\ Dvali,
 Phys.\ Lett.\ B436 (1998) 257.
\bibitem{burt}A.\ Lukas, B.A.\ Ovrut, K.S.\ Stelle, D.\ Waldram,
  Phys.\ Rev.\ D59 (1999) 086001, Nucl.\ Phys.\ B552 (1999) 246.
\bibitem{pierre1}P.\ Bin\'{e}truy, C.\ Deffayet, D.\ Langlois,
  Nucl.\ Phys.\ B565 (2000) 269.
\bibitem{RS1}L.\ Randall, R.\ Sundrum,
  Phys.\ Rev.\ Lett.\ 83 (1999) 3370.
\bibitem{keith}K.R.\ Dienes, E.\ Dudas, T.\ Gherghetta,
  Nucl.\ Phys.\ B567 (2000) 111. 
\bibitem{klaus}K.\ Behrndt, M.\ Cveti\v{c},
  Phys.\ Lett.\ B475 (2000) 253. 
\bibitem{gata}J.\ Garriga, T.\ Tanaka,
  Phys.\ Rev.\ Lett.\ 84 (2000) 2778.
\bibitem{dimi}R.\ Dick, D.\ Mikulovi\'{c},
  Phys.\ Lett.\ B476 (2000) 363.
\bibitem{wolf1}W.\ M\"uck, K.S.\ Viswanathan,
 I.V.\ Volovich, hep-th/0002132.
\bibitem{wolf2}I.Ya.\ Aref'eva, M.G.\ Ivanov,
 W.\ M\"uck, K.S.\ Viswanathan,
 I.V.\ Volovich, hep-th/0004114.
\bibitem{ioannis}I.\ Giannakis, H.-c.\ Ren,
  hep-th/0007053.
\bibitem{nihei}T.\ Nihei, 
  Phys.\ Lett.\ B465 (1999) 81.
\bibitem{groj1}J.M.\ Cline, C.\ Grojean, G.\ Servant,
  Phys.\ Rev.\ Lett.\ 83 (1999) 4245.
\bibitem{kraus}P.\ Kraus, JHEP 9912 (1999) 011.
\bibitem{yuri}A.\ Yu.\ Neronov, Phys.\ Lett.\ B472 (2000) 273.
\bibitem{pierre2}P.\ Bin\'{e}truy, C.\ Deffayet, U.\ Ellwanger, 
  D.\ Langlois, 
  Phys.\ Lett.\ B477 (2000) 285.
\bibitem{vollick}D.N.\ Vollick, hep-th/9911181.
\bibitem{ida}D.\ Ida, gr-qc/9912002.
\bibitem{cvetic2}M.\ Cveti\v{c}, J.\ Wang, 
  Phys.\ Rev.\ D61 (2000) 124020.
\bibitem{cline}J.M.\ Cline, hep-ph/0001285.
\bibitem{holdom}H.\ Collins, B.\ Holdom, hep-ph/0003173v2.
\bibitem{deruelle}N.\ Deruelle, T.\ Dole\v{z}el,
  gr-qc/0004021.
\bibitem{tye}H.\ Stoica, S.-H.H.\ Tye, I.\ Wasserman,
 Phys.\ Lett.\ B482 (2000) 205.
\bibitem{lanczos}K.\ Lanczos, 
  Annalen Phys.\ (4.\ Serie) 74 (1924) 518.
\bibitem{israel}W.\ Israel, Nuovo Cim.\ X 44B (1966) 1,
 (E) X 48B (1967) 463.
\bibitem{reall}H.A.\ Chamblin, H.S.\ Reall, Nucl.\ Phys.\ 
  B562 (1999) 133.

\end{thebibliography}
\end{document}